\documentclass[twocolumn,tighten]{aastex63}
\usepackage{amssymb, amsmath,framed}
\usepackage{gensymb}
\usepackage{savesym}
\savesymbol{tablenum}
\usepackage{siunitx}
\restoresymbol{SIX}{tablenum}
\usepackage{rotating}

\usepackage{afterpage}

\usepackage{bm}
\expandafter\ifx\csname package@font\endcsname\relax\else
 \expandafter\expandafter
 \expandafter\usepackage
 \expandafter\expandafter
 \expandafter{\csname package@font\endcsname}

\fi
\def\bq{\begin{equation}}
\def\eq{\end{equation}}
\def\bqy{\begin{eqnarray}}
\def\eqy{\end{eqnarray}}
\def\p{\partial}

\def\p{\partial}

\def\R{\mathbb{R}}

\begin{document}
\title{\large{Project Lyra: Another Possible Trajectory to 1I/'Oumuamua}}

\correspondingauthor{Adam Hibberd}
\email{adam.hibberd@i4is.org}

\author{Adam Hibberd}
\affiliation{Initiative for Interstellar Studies (i4is) 27/29 South Lambeth Road London, SW8 1SZ United Kingdom}

\begin{abstract}
 
 The first interstellar object to be discovered, 1I/'Oumuamua, exhibited various unusual properties as it was tracked on its passage through the inner solar system in 2017/2018. In terms of the potential scientific return, a spacecraft mission to intercept and study it in situ would be invaluable. As an extension to previous Project Lyra studies, this paper elaborates an alternative mission to 1I/'Oumuamua, this time also requiring a Jupiter Oberth Manoeuvre (JOM) to accelerate the spacecraft towards its destination. The difference is in the combination of planetary flybys exploited to get to Jupiter, which includes a Mars encounter before proceeding to Jupiter. The trajectory identified is inferior to previous finds in terms of higher  $\Delta V$ requirement (15.6 \si{km.s^{-1}}), longer flight duration (29 years) and less mission preparation time (launch 2026), however it benefits from a feature absent from previous JOM candidates, in that there is little or no $\Delta V$ en route to Jupiter (i.e. a free ride) which means the spacecraft need not carry a liquid propellant stage. This is marginally offset by the higher $\Delta V$ needed at Jupiter, requiring either 2 or 3 staged solid rocket motors. As an example, a Falcon Heavy Expendable with a CASTOR 30B booster followed by a STAR 48B can deliver 102kg to 1I/'Oumuamua by the year 2059.  Other scenarios with shorter flight durations and higher payload masses are possible.
\\
 
\end{abstract}

\section{Introduction} \label{SecIntro}

1I/'Oumuamua is the first interstellar object (ISO) discovered passing through the inner solar system and unequivocally identified as such, with orbital eccentricity $e = 1.2$ (i.e. $e > 1$). Many origin theories exist to resolve the nature of this mysterious extrasolar object, each with their own merits and degrees of success at explaining the unusual observational data \citep{Flekky2019,Seligman2020,Jackson2021,Desch2021,Bialy2018,Raymond2018}.\\

As it is currently out of sight of existing Earth and space-based telescopes, its true nature may never be understood - unless a spacecraft mission to study it in situ can be realised. Various previous Project Lyra studies have identified a Solar Oberth Manoeuvre (SOM) as a viable option \citep{Hein2019b,HIBBERD2021594,HIBBERD2021,HIBBERD2020} and more recent analysis indicates a Jupiter Oberth Manoeuvre (JOM), with the route to Jupiter exploiting the VEEGA sequence of planetary encounters \citep{https://doi.org/10.48550/arxiv.2201.04240}.\\

The VEEGA JOM candidate propounded in \cite{https://doi.org/10.48550/arxiv.2201.04240} has the disadvantage of requiring several applications of $\Delta V$ along the trajectory to Jupiter and therefore would need a liquid propellant stage to generate this thrust, in addition to the 2 solid propellant stages required to deliver the JOM $\Delta V$.\\

The trajectory proposed in this paper does not suffer this drawback and so enables a higher mass to be brought to bear at Jupiter for the JOM. The mission is abbreviated as V-E-DSM-E-M-J.\\ 
\begin{figure}[h]
\centering
\includegraphics[scale=0.30]{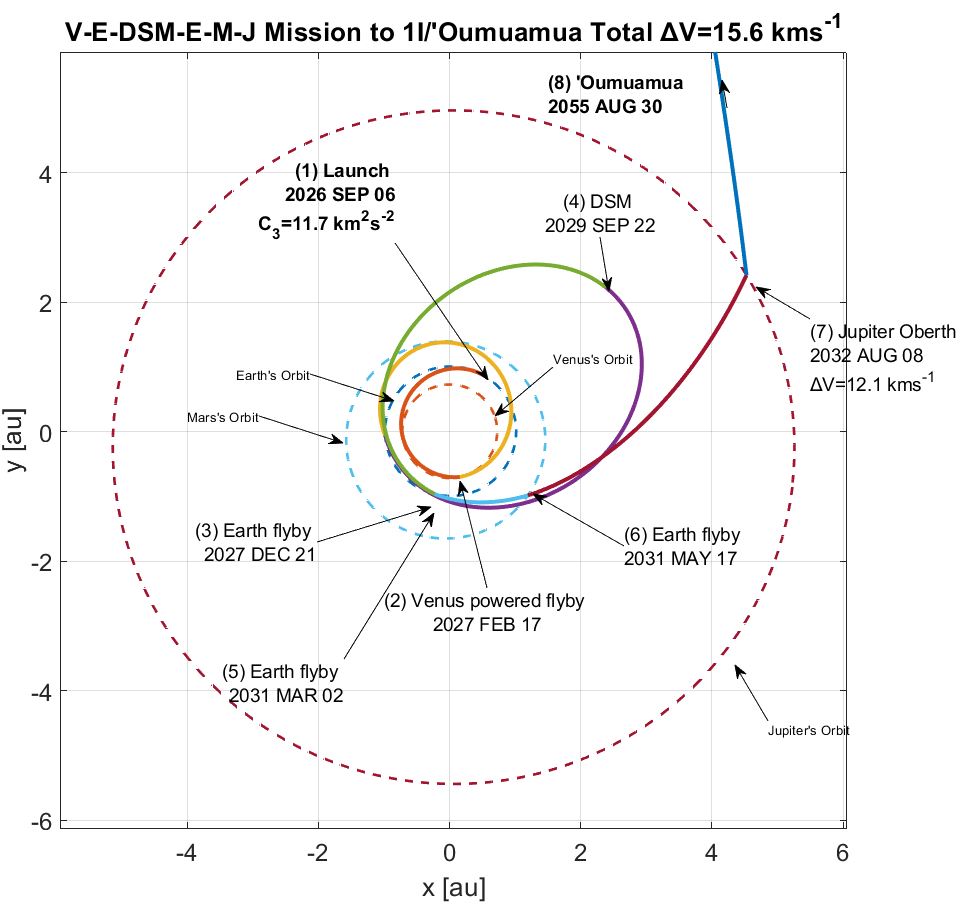}
\caption{Trajectory, V-E-DSM-E-M-J}
\label{fig:Traj}
\end{figure}
\afterpage{
\begin{sidewaystable*}[]
\centering
\vspace*{+8cm}
\caption{Possible Combinations of Solid Stages for the JOM}
\label{tab:STS_combi}

\begin{tabular}{|c|c|c|c|c|c|c|}
\hline
\textbf{Falcon Heavy/SLS   Block 1B/Block 2} &
  \textbf{Mtot \textless 17000kg} &
   &
   &
   &
   &
   \\ \hline
Stage 1 &
  Stage 2 &
  Stage 3 &
  $\Delta V_{JOM}$ &
  Total Mass &
  Payload &
  Arrival Year \\ \hline
\multicolumn{1}{|l|}{} &
  \multicolumn{1}{l|}{} &
  \multicolumn{1}{l|}{} &
  (\si{km.s^{-1}}) &
  JOM (kg) &
  Mass (kg) &
   \\ \hline
STAR 75 &
  STAR 48B &
  N/A &
  11.2 &
  10232 &
  27 &
  2059 \\ \hline
CASTOR 30B &
  STAR 48B &
  N/A &
  11.2 &
  16210 &
  102.16 &
  2059 \\ \hline
CASTOR 30B &
  STAR 48B &
  N/A &
  12.1 &
  16146 &
  38 &
  2055 \\ \hline
STAR 75 &
  STAR 63F &
  STAR 48B &
  11.2 &
  14902 &
  107 &
  2059 \\ \hline
 &
   &
   &
   &
   &
   &
   \\ \hline
\textbf{SLS Block 1B/Block 2} &
  \textbf{17000kg \textless Mtot \textless 34000kg} &
   &
   &
   &
   &
   \\ \hline
Stage 1 &
  Stage 2 &
  Stage 3 &
  $\Delta V_{JOM}$ &
  Total Mass &
  Payload &
  Arrival Year \\ \hline
\multicolumn{1}{|l|}{} &
  \multicolumn{1}{l|}{} &
  \multicolumn{1}{l|}{} &
  (\si{km.s^{-1}}) &
  JOM (kg) &
  Mass (kg) &
   \\ \hline
CASTOR 30XL &
  STAR 48B &
  N/A &
  12.1 &
  28671 &
  128 &
  2055 \\ \hline
CASTOR 30XL &
  STAR 63F &
  N/A &
  12.1 &
  31039 &
  43 &
  2055 \\ \hline
CASTOR 30XL &
  STAR 63F &
  STAR 48B &
  12.1 &
  33361 &
  228 &
  2055 \\ \hline
CASTOR 30B &
  STAR 75 &
  STAR 48B &
  12.1 &
  24301 &
  125 &
  2055 \\ \hline
CASTOR 30B &
  STAR 63F &
  STAR 48B &
  12.1 &
  20806 &
  108 &
  2055 \\ \hline
CASTOR 30XL &
  STAR 48B &
  N/A &
  13.2 &
  28586 &
  43 &
  2052 \\ \hline
CASTOR 30XL &
  ORION 50XL &
  STAR 48B &
  13.2 &
  32949 &
  99 &
  2052 \\ \hline
CASTOR 30XL &
  STAR 63F &
  STAR 48B &
  13.2 &
  33244 &
  111 &
  2052 \\ \hline
CASTOR 30XL &
  STAR 63F &
  STAR 48B &
  11.2 &
  33501 &
  368 &
  2059 \\ \hline
CASTOR 30B &
  STAR 75 &
  STAR 48B &
  11.2 &
  24400 &
  224 &
  2059 \\ \hline
CASTOR 30XL &
  STAR 48B &
  N/A &
  11.2 &
  28773 &
  230 &
  2059 \\ \hline
 &
   &
   &
   &
   &
   &
   \\ \hline
\textbf{SLS Block 2} &
  \textbf{34000kg \textless Mtot \textless 38000kg} &
   &
   &
   &
   &
   \\ \hline
Stage 1 &
  Stage 2 &
  Stage 3 &
  $\Delta V_{JOM}$ &
  Total Mass &
  Payload &
  Arrival Year \\ \hline
\multicolumn{1}{|l|}{} &
  \multicolumn{1}{l|}{} &
  \multicolumn{1}{l|}{} &
  (\si{km.s^{-1}}) &
  JOM (kg) &
  Mass (kg) &
   \\ \hline
CASTOR 30XL &
  STAR 75 &
  STAR 48B &
  11.2 &
  37000 &
  388 &
  2059 \\ \hline
CASTOR 30XL &
  STAR 75 &
  STAR 48B &
  12.1 &
  36853 &
  241 &
  2055 \\ \hline
CASTOR 30XL &
  STAR 75 &
  STAR 48B &
  13.2 &
  36730 &
  119 &
  2052 \\ \hline
\end{tabular}
\end{sidewaystable*}[]
}
\section{Method}

\begin{table*}[]
\centering
\caption{Mission V-E-DSM-E-M-J with 29 Years Flight Duration Constraint}
\label{tab:Traj_info}
\begin{tabular}{|c|c|c|c|c|c|c|c|}
\hline
\textbf{Num} & \textbf{Planet} & \textbf{Time} & \textbf{Arrival speed} & \textbf{Departure speed} & \textbf{$\Delta V$} & \textbf{Cumulative $\Delta V$} & \textbf{Periapsis} \\
  &              &               &  (\si{km.s^{-1}})  &  (\si{km.s^{-1}})   &   (\si{km.s^{-1}})  &  (\si{km.s^{-1}})  &  \textbf{alt.} (km) \\ \hline
1 & Earth        & 2026 SEP 06 & 0.0000  & 3.4199  & 3.4199  & 3.4199  & N/A     \\ \hline
2 & Venus        & 2027 FEB 17 & 6.6646  & 6.6629  & 0.0013  & 3.4212  & 11987.3 \\ \hline
3 & Earth        & 2027 DEC 21 & 10.0464 & 10.0464 & 0.0000  & 3.4212  & 352.2   \\ \hline
4 & DSM @ 3.2 au & 2029 SEP 22 & 11.0568 & 11.0799 & 0.0232  & 3.4444  & N/A     \\ \hline
5 & Earth        & 2031 MAR 02 & 9.8700  & 9.8700  & 0.0000  & 3.4444  & 4966.1  \\ \hline
6 & Mars         & 2031 MAY 17 & 18.9898 & 18.9898 & 0.0000  & 3.4444  & 200     \\ \hline
7 & Jupiter      & 2032 AUG 08 & 12.0498 & 37.4067 & 12.1175 & 15.5619 & 58969.4 \\ \hline
8 & 1I/'Oumuamua     & 2055 AUG 30 & 17.5022 & 17.5022 & 0.0000  & 15.5619 & N/A     \\ \hline
\end{tabular}
\end{table*}
Optimum Interplanetary Trajectory Software (OITS) was used to generate the solution trajectories investigated herein. The theory is summarised in \cite{HIBBERD2021584} and further detailed in \cite{OITS_info}. High thrust propulsion is assumed, and chemical propellant rockets are chosen. A minimum periapsis altitude of 200 km at each planetary encounter is adopted.\\

As in previous studies, two Non-Linear Programming (NLP) optimizers were used, NOMAD \citep{LeDigabel2011} and MIDACO \citep{Schlueter_et_al_2009,Schlueter_Gerdts_2010,Schlueter_et_al_2013}.\\
\section{Results}
\begin{figure}[h]
\hspace{-1cm}
\centering
\includegraphics[scale=0.31]{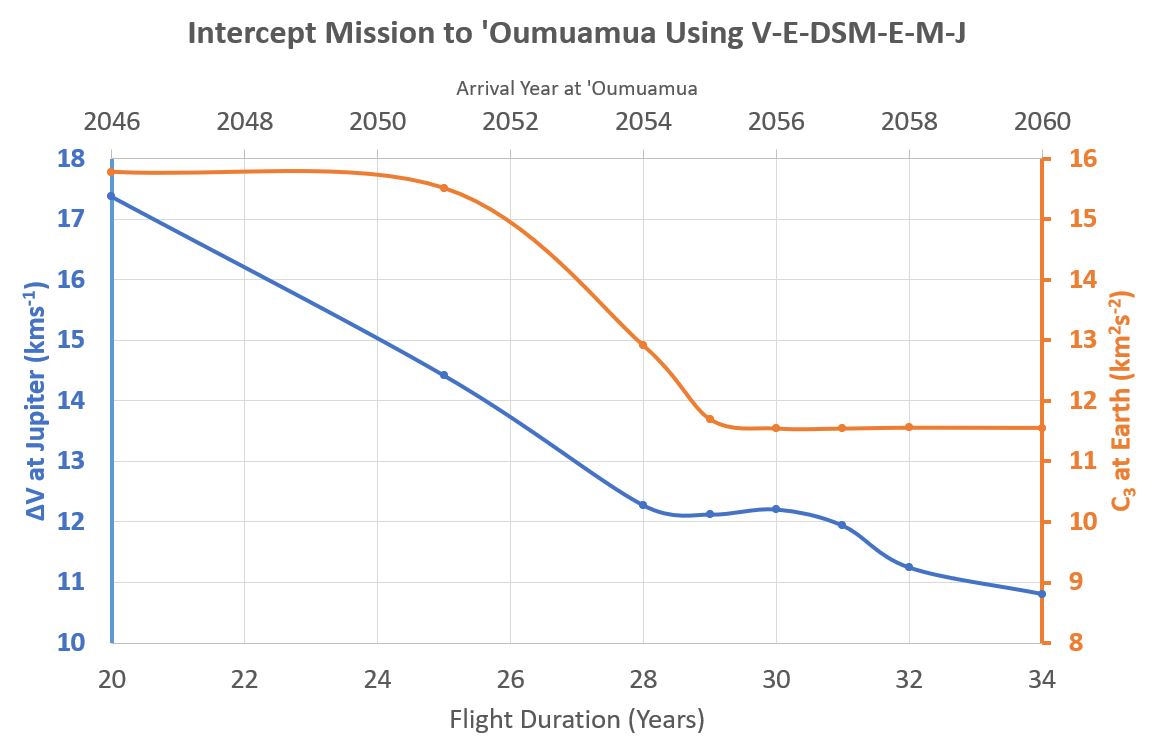}
\caption{ $\Delta$V vs Flight Duration, V-E-DSM-E-M-J }
\label{fig:Traj_Data}
\end{figure}

\begin{figure}[h]
\hspace{-0.71cm}
\centering
\includegraphics[scale=0.31]{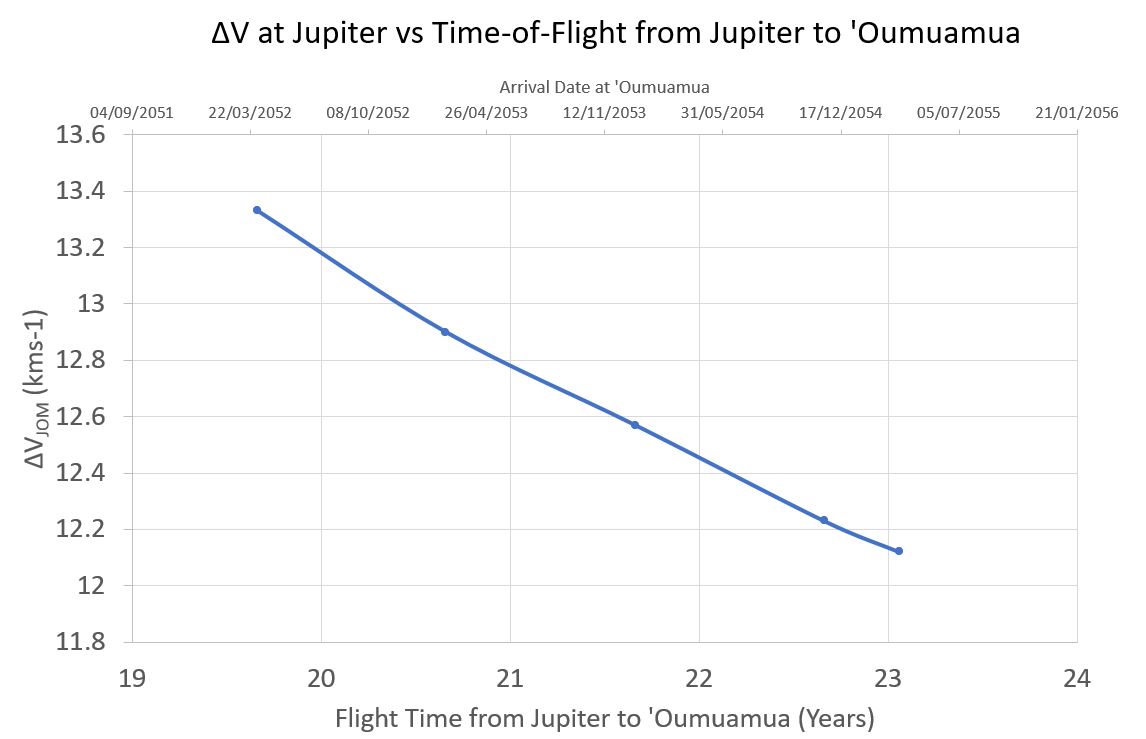}
\caption{ $\Delta V_{JOM}$ vs Jupiter to 1I/'Oumuamua Time-of-Flight}
\label{fig:Traj_Data2}
\end{figure}

\begin{table*}[]
\centering
\caption{Solid Rocket Stages Considered for JOM $\Delta V$}
\label{tab:SRS_info}
\begin{tabular}{|c|c|c|c|}
\hline
\textbf{Solid Rocket Stage} & \textbf{Total Mass (\si{kg})} & \textbf{Dry Mass (\si{kg})} & \textbf{Exhaust Velocity (\si{km.s^{-1}})} \\ \hline
STAR 48B    & 2137  & 124  & 2.8028 \\ \hline
STAR 63F    & 4590  & 326  & 2.9106 \\ \hline
STAR 75     & 8068  & 565  & 2.8224 \\ \hline
ORION 50XL  & 4306  & 367  & 2.8647 \\ \hline
CASTOR 30B  & 13971 & 1000 & 2.9649 \\ \hline
CASTOR 30XL & 26406 & 1392 & 2.8866 \\ \hline
\end{tabular}
\end{table*}
\begin{table*}[]
\centering
\caption{Payload Mass capability for 3 Launch Vehicles}
\label{tab:Launcher_info}
\begin{tabular}{|c|c|}
\hline
\textbf{Launch Vehicle} & \textbf{Mass to $C_3 = 11.7\si{km^2.s^{-2}}$} \\ \hline
Falcon Heavy            & 17000                                         \\ \hline
SLS Block 1B            & 34000                                         \\ \hline
SLS Block 2             & 38000                                         \\ \hline
\end{tabular}
\end{table*}

Table \ref{tab:STS_combi} provides the payload masses achievable using the V-E-DSM-E-M-J combination of gravitational assists (GA). The solution trajectory is shown in Figure \ref{fig:Traj} and the numerical data is provided in Table \ref{tab:Traj_info}. An animation of this can be found at \cite{CiteTrajAnim2}. This trajectory is included in the optimal missions to Jupiter list in \cite{doi:10.2514/2.3650}). From Table \ref{tab:Traj_info} we see the Earth hyperbolic excess for this 29 year mission is 3.4199 \si{km.s^{-1}} (row 1) giving the Characteristic Energy at Earth of $C_{3} = 3.4199^{2} = 11.70 \si{km^2.s^{-2}}$. Virtually the only $\Delta V$ remaining along the trajectory is that for the JOM and is $\Delta V_{JOM}=12.1 \si{km.s^{-1}}$. \\

It is clearly of relevance to understand how these two parameters of $C_{3}$ and $\Delta V_{JOM}$ evolve with flight duration constraint. To this end, refer to Figure \ref{fig:Traj_Data} which provides this information for durations between 20 and 34 years, corresponding to arrival years 2046 to 2060. The hump in the $\Delta V_{JOM}$ values for flight times from 28 years to 32 years appears to be a genuine phenomenon and not a consequence of the NLP solvers finding local minima solutions over this interval.\\

From Figure \ref{fig:Traj_Data} we see a consequence of this hump is that there is a trough in $\Delta V_{JOM}$ at around 29 years and indeed the $C_{3}$ reaches approximately the same value as for longer durations. This supports the decision to adopt a nominal flight duration of 29 years for this trajectory profile.\\

We can observe in Figure \ref{fig:Traj_Data} the effect of either contracting or expanding the overall flight duration on the optimal $\Delta V$ and $C_{3}$ values, where optimality is defined as the minimium of the sum of the hyperbolic excess at Earth with that required en route, including the JOM. However it is also of interest to consider the effect on the $\Delta V_{JOM}$ of only modifying the Time-of-Flight from Jupiter to 1I/'Oumuamua whilst constraining the preceding $\Delta V$s, as well as the $C_3$ value at Earth, to those provided in Table \ref{tab:Traj_info} .\\ 

Figure \ref{fig:Traj_Data2} is a graph providing precisely this information and as can be observed, all other $\Delta V$s being equal, for arrival years at 1I/'Oumuamua from 2052 to 2055, the $\Delta V_{JOM}$ drops from 13.3 \si{km.s^{-1}} down to 12.1 \si{km.s^{-1}}.\\

We can now address the question of how a mission might exploit this trajectory profile in practice, i.e. which combination of launch vehicle and spacecraft propulsion system might be needed. From Table \ref{tab:Traj_info}, we see the required $C_3$ is 11.7 \si{km^2.s^{-2}}. A list of approximate payload capacities for three launch vehicles, SpaceX Falcon Heavy Expendable, NASA SLS Block 1B and SLS Block 2, is provided in Table \ref{tab:Launcher_info}.\\

Assuming that the JOM kick is delivered by a solid rocket, the level of $\Delta V_{JOM}$ required (i.e. $>$ 11 \si{km.s^{-1}}), would necessitate at least two stages and ideally three. Table \ref{tab:SRS_info} provides the list of candidate solid booster stages which shall be considered here.\\

A list of viable combinations of these solid booster stages, numbering either 2 or 3, is provided in Table \ref{tab:STS_combi}, it should be noted that this is not an exhaustive list.\\

Referring to Table \ref{tab:STS_combi}, we see that a Falcon Heavy with a pairing of CASTOR 30B followed by a STAR 48B can deliver 102kg to 1I/'Oumuamua by the year 2059. With a suitably sized fairing, an SLS Bock 1B can achieve 128kg by 2055, using a CASTOR 30XL, followed by a STAR 48B. The highest mass possible however is for an SLS Block 2 which using a CASTOR 30XL, a STAR 75 and then a STAR 48B can reach 1I/'Oumuamua by 2059 with a payload of 388kg. For the latter, it may be unlikely that the Fairing capacity would be sufficient to contain all the stages and the payload.

\section{Discussion}

The key feature of the V-E-DSM-E-M-J trajectory investigated above, making it a noteworthy and attractive proposition for a mission, is the absence of burns needed between Earth launch and the JOM. In addition the approach velocity of the spacecraft relative to 1I/'Oumuamua is 17.5 \si{km.s^{-1}}, lower than that achieved in previous Project Lyra studies. However there are three disadvantages of this trajectory in that, first, the $\Delta V$ required at the JOM is high ($>$ 11 \si{km.s^{-1}}); second, at the time of writing, the launch date is in only 4 years hence (interplanetary missions can take 5-10 years from inception to launch) and third, the flight duration is rather extended (i.e. $>$ 26 years as opposed to 22 years with a SOM, \cite{HIBBERD2020}) .

\section{Conclusion}
In this paper a mission to 1I/'Oumuamua using a series of unpowered GAs followed by a powered flyby of Jupiter - a JOM - is investigated. The route studied is abbreviated as V-E-DSM-E-M-J, and its superiority over previously studied Project Lyra trajectories is down to there being essentially no thrust needed en route to Jupiter, meaning that no liquid propellant stage is required which would otherwise take up much needed capacity for the JOM. Assuming a 29 year flight duration selection, the total $\Delta V$ is 15.6 \si{km.s^{-1}}. Masses of typically around 120 kg, and theoretically up to 388 kg, can be achieved depending on particular launcher and solid propellant stages exploited. 

\bibliographystyle{aasjournal}
\bibliography{library,library_Adam_Hibberd,Hein_ISO_Modified_by_Adam_Hibberd}

\end{document}